\documentclass{ws-ijmpe}
\usepackage{graphicx}
\begin{document}

\markboth{Claude Pruneau}{Methods for Jet Studies with Three-Particle Correlations }

\catchline{}{}{}{}{}

\title{Methods for Jet Studies with Three-Particle Correlations.}

\author{\footnotesize Claude Pruneau}

\address{ Physics and Astronomy Department, Wayne State University, \\
666 West Hancock, Detroit, Michigan, 48152 USA\\
pruneau@physics.waynne.edu }

\maketitle

\begin{history}
\received{(received date)}
\revised{(revised date)}
\end{history}

\abstract
{
We present a method based on three-particle azimuthal correlation cumulants for the study of the interaction of jets with the medium produced in heavy ion collisions at RHIC and LHC where jets cannot be reconstructed on an event-by-event basis with conventional jet finding algorithms.  The method is specifically designed to distinguish a range of jet interaction mechanisms such as Mach cone emission, gluon Cerenkov emission, jet scattering, and jet broadening. We discuss how anisotropic flow background contibutions of second order (e.g. $v_2^2$) are suppressed in three-particle azimuthal correlation cumulants, and discussed specific model representations of di-jets, away-side scattering, and Mach cone emission.
}

\section{Introduction}

The observation of a dip at $180^o$ in flow subtracted two-particle azimuthal correlations observed in
 Au + Au collisions \cite{Star05} has recently generated quite a bit of interest. Stoecker \cite{Stoecker05} and independently Shuryak {\em et al.} \cite{Shuryak05}  suggested the observed dip might be an indicator of the production of away-side parton induced wake field or Mach cone, with the width of the cone determined by the sound velocity in the produced medium.  The dip might however also result from other processes such as large angle gluon radiation \cite{Vitev05},  jets deflection by radial flow, or Cerenkov gluon radiation \cite{Majumber06}.  While discrimination of these production mechanisms is not possible with two-particle 
correlations, it might be achieved with three particle correlations. Preliminary such three particle analyses presented at QM05 and elsewhere by STAR \cite{Ulery05},  and PHENIX \cite{Ajitanand06} arise as rather challenging tasks. Observation of three-particle correlations require large datasets. One must also account for the fact that measured three-particle densities consist of a superposition of correlated three particle signals,
and combinatorial terms involving only two correlated or three uncorrelated particles. Finally, the interpretation of the data is also complicated by finite anisotropic flow, and momentum conservation effects \cite{Borghini06}.

We present and discuss the merits of an analysis technique based on cumulants. The definition of cumulants, measurement method, and key properties are discussed in Section 2. Section 3 includes a brief discussion of correlation shapes expected from simple models of in-vacuum jets, deflected jets, conical emission, and jet-flow. 

Some of these topics were already discussed in \cite{Pruneau06}. Also note that alternative analysis techniques were discussed in the recent literature \cite{Ajitanand06,Ulery06}.

\section{Cumulant Definition and Properties}

The cumulant method makes no assumption about the background or shape of the signal. Cumulants are defined as statistical measures of the degree of correlation between measured particles. For three particles, the definition reads:
\begin{equation}
\begin{array}{c}
 C_3 (\varphi _i ,\varphi _j ,\varphi _k ) = \rho _3 (\varphi _i ,\varphi _j ,\varphi _k ) - \rho _2 (\varphi _i ,\varphi _j )\rho _1 (\varphi _k ) - \rho _2 (\varphi _i ,\varphi _k )\rho _1 (\varphi _j ) \\ - \rho _2 (\varphi _j ,\varphi _k )\rho _1 (\varphi _i ) 
  + 2\rho _1 (\varphi _i )\rho _1 (\varphi _j )\rho _1 (\varphi _k ) \\ 
 \end{array}
\end{equation}	
where $\rho _3 (\varphi _i ,\varphi _j ,\varphi _k ) = dN/d\varphi _i d\varphi _j d\varphi _k$, $\rho _2 (\varphi _i ,\varphi _j ) = dN/d\varphi _i d\varphi _j$, and $\rho _1 (\varphi _i ) = dN/d\varphi _i$, are respectively three, two, and single particle densities (normalized per event), measured for selected particles identified with labels i, j, and k. The definition holds whether one considers identical particles, or different species, or whether the integrated kinematic ranges are the same, overlap, or are different for the three particles measured.  

By construction, cumulants are non-positive definite quantities, and indicate the degree to which particles are correlated relative to Poisson processes: positive values indicate regions of phase space where particles are likely to be founded together, whereas negative values signal regions where particles are unlikely compared to uncorrelated particle production (Poisson). It is straightforward to show that the cumulant of a sum of independent processes is equal to the sum of the cumulant of each the processes thereby enabling modeling of particle production in terms of separate components, such as jets, and background particles. Note however that such a separation in a finite size system leads to sampling biases: cumulants are non-zero even in the absence of correlations. 

The definition Eq. 1 is of limited interest for the study of the physical processes involved in the particle production. More relevant is a formulation in terms of relative angles. With three particles, there are two such independent angles we here choose as $\Delta \varphi _{ij}  = \varphi _i  - \varphi _j$ and $\Delta \varphi _{ik}  = \varphi _i  - \varphi _k$. The cumulant in terms of these relative angles can be obtained by summing Eq. 1 over all phase space.
\begin{equation}
C_3 (\Delta \varphi _{ij} ,\Delta \varphi _{ik} ) = \int {C_3 (\varphi _i ,\varphi _j ,\varphi _k )} \delta (\Delta \varphi _{ij}  - \varphi _i  + \varphi _j )\delta (\Delta \varphi _{ik}  - \varphi _i  + \varphi _k )d\varphi _i d\varphi _j d\varphi _k 
\end{equation}
The cumulants must be corrected for finite detection efficiencies, and averaged over collision centralities in heavy ion collisions. While this can be accomplished in a variety of ways, we advocate a simple technique based on the hypothesis that the efficiencies for simultaneously detecting multiple particles can be factorized as a product of the efficiencies for detecting each of the particles. Labeling the uncorrected cumulant as "raw", the efficiency corrected cumulant may be obtained as follows:
	\begin{equation}
C_3 (\Delta \varphi _{ij} ,\Delta \varphi _{ik} ) = \frac{{\overline {\rho _1^{(raw)} (i)} }}{{\overline {\varepsilon (i)} }}\frac{{\overline {\rho _1^{(raw)} (j)} }}{{\overline {\varepsilon (j)} }}\frac{{\overline {\rho _1^{(raw)} (k)} }}{{\overline {\varepsilon (k)} }}\frac{{C_{_3 }^{(raw)} (\Delta \varphi _{ij} ,\Delta \varphi _{ik} )}}{{\rho _1 \rho _1 \rho _{_1 }^{(raw)} (\Delta \varphi _{ij} ,\Delta \varphi _{ik} )}}
\end{equation}

where the overbar indicate averaging over all azimuths, and $\rho _{_1 }\rho _{_1 }\rho _{_1 }^{(raw)}(\Delta \varphi _{ij} ,\Delta \varphi _{ik} )$ is calculated as follows:
\begin{equation}
\begin{array}{c}
\rho _{_1 }\rho _{_1 }\rho _{_1 }^{(raw)}(\Delta \varphi _{ij},\Delta\varphi _{ik} )= \int {\rho _1^{(raw)}(\varphi _i)\rho _1^{(raw)} (\varphi _j )\rho _1^{(raw)} (\varphi _k )} \\ 
\times \delta (\Delta \varphi _{ij}-\varphi _i  + \varphi _j )\delta (\Delta \varphi _{ik}-\varphi _i  + \varphi _k )d\varphi _i d\varphi _j d\varphi _k  \\ 
\end{array}
\end{equation}
Fig. 1 shows examples of three-particle density, combinatorial terms and 3-cumulants for 10-30\% Au + Au collisions. 

\begin{figure}[htb]
\begin{center}
\includegraphics[width=\linewidth,height=2.in]{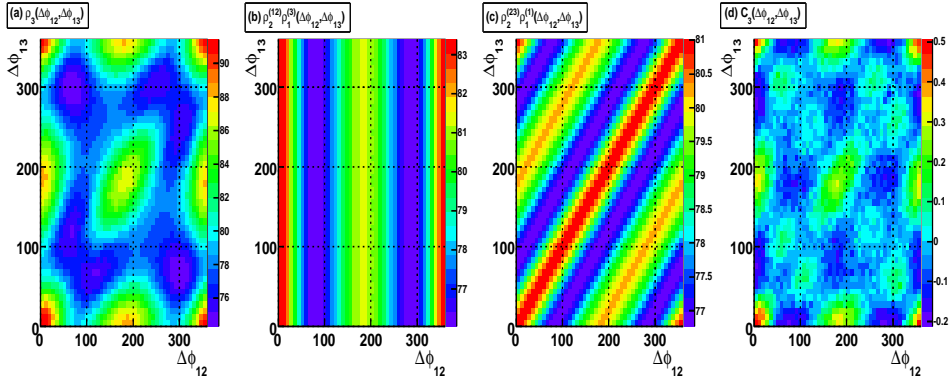} \end{center}
\caption[]{Examples of (a) 3-particle density, (b,c) combinatorial terms, and (d) 3-cumulant obtained for Au + Au collisions at RHIC. Angles expressed in degrees. See text for definitions.}
\label{fig1}
\end{figure}

\section{Flow, Jets, and Conical Emission Discrimination}

Processes such as jet production, deflected jet production, and conical emission lead to distinct kinematical signature in $
C_3 (\Delta \varphi _{ij} ,\Delta \varphi _{ik} )$. In \cite{Pruneau06}, we modeled jet production within a Gaussian approximation, and found the away-side jet 3-cumulant component is proportional to $\exp \left( { - \left( {\Delta \varphi _{12}  - \pi } \right)^2 /2\sigma _{12}^2  - \left( {\Delta \varphi _{13}  - \pi } \right)^2 /2\sigma _{13}^2  - \left( {\Delta \varphi _{23}} \right)^2 /2\sigma _{23}^2 } \right)$ with an amplitude determined by the number of jets per event, and the number of associated particles per jet. The widths $\sigma _{12} $ and $\sigma _{13}$, somewhat larger than $\sigma _{23} $ for in vacuum jets, are expected to significantly increase for in medium deflected jets while $\sigma _{23} $ which measures the "intrinsic" width of the jet changes only modestly. Mach cone emission is, in principle, easily distinguishable from jet production as it leads to four jacobian peaks in $C_3$  at $\Delta \varphi _{ij} ,\Delta \varphi _{ik} = {\pi  \pm \theta _M ,\pi  \pm \theta _M }$ where $\theta _M $ is determined by the sound velocity in the produced medium . The shape and width of the peaks shall depend on the details of the production process and may be influenced by radial flow effects. Fig. 2 illustrates the sensitivity of the technique based on a simulation involving near side jets with a width of 0.2 radian and conical emission at $\theta _M $ with 0.4 radian width. It shows the 3-cumulants and projections along the main and alternate diagonals in slices of 40 degrees. Plots were generated with a sample of 8 million events. Particle production was carried out with Poisson generators such that the event multiplicity ranges from 300 to 600, with an average of 0.23 jets per event, and averaged associated multiplicities of one for the jet tag particle "1" and two for associates "2" and "3", in the near side jet and cone. 

The observation of conical emission and study of jet structure is complicated by the presence of irreducible flow components in the 3-cumulant. Assuming "background" particle emission is correlated to the reaction plane orientation and can be described with a Fourier decomposition e.g. $ \propto 1 + 2\sum\limits_n {v_n \cos \left( {n\left( {\varphi _i  - \psi } \right)} \right)} $ where $v_n$ and $\psi$ are respectively flow coefficients and reaction plane angle, one finds the 3-cumulant shall contain contributions involving "off-diagonal" terms of the form: 
\begin{equation}
\sum\limits_{p,m,n}^{} {v_p (i)v_m (j)v_n (k)}  \times \left[ \begin{array}{l}
 \delta _{p,m + n} \cos \left( {p\varphi _i  - m\varphi _j  - n\varphi _k } \right) \\ 
  + \delta _{m,p + n} \cos \left( { - p\varphi _i  + m\varphi _j  - n\varphi _k } \right) \\ 
  + \delta _{n,m + k} \cos \left( { - p\varphi _i  - m\varphi _j  + n\varphi _k } \right) \\ 
 \end{array} \right]
\end{equation}
Based on existing measurements of flow coefficients (see for instance \cite{YutinBai06}), we expect such contributions to be dominated by terms of order $v_2v_2v_4$ as indeed found in Fig. 1 (See \cite{Pruneau06} for details). 

The extraction of conical emission signal may be further complicated by correlated jet emission with the reaction plane. Indeed, one expects jet quenching or attenuation shall depend on parton path length in the medium. Given the finite spatial eccentricity of the medium produced in mid-central collisions, this implies jet emission can be correlated with the reaction plane orientation. Again using a Gaussian formulation for the jet profile, one finds that the 3-cumulant should contain jet-jet and jet-background terms of the form:
\begin{equation}
v_2 (jet)v_2 (bckg)\exp \left( { - \Delta \varphi _{ij}^2 /2\sigma _{ij}^2 } \right)\cos \left( {2\left( {a_i \varphi _i  + a_j \varphi _j  - \varphi _k } \right)} \right)
\end{equation}
where the coefficients $a_i$ and $a_j$ are determined by the relative widths of the jets.  Fig. 3 illustrates the shape of 3-cumulant resulting from such terms where we used $a_i = a_j = 0.5$. Clearly, the reaction plane correlation produces an away-side narrow modulation, which should not be confused with three-particle jet contributions given its characteristic cosine dependence.

\begin{figure}[htb]
\mbox{
\begin{minipage}{0.5\linewidth}
\begin{center}\includegraphics[width=3.1in,height=3.8in]{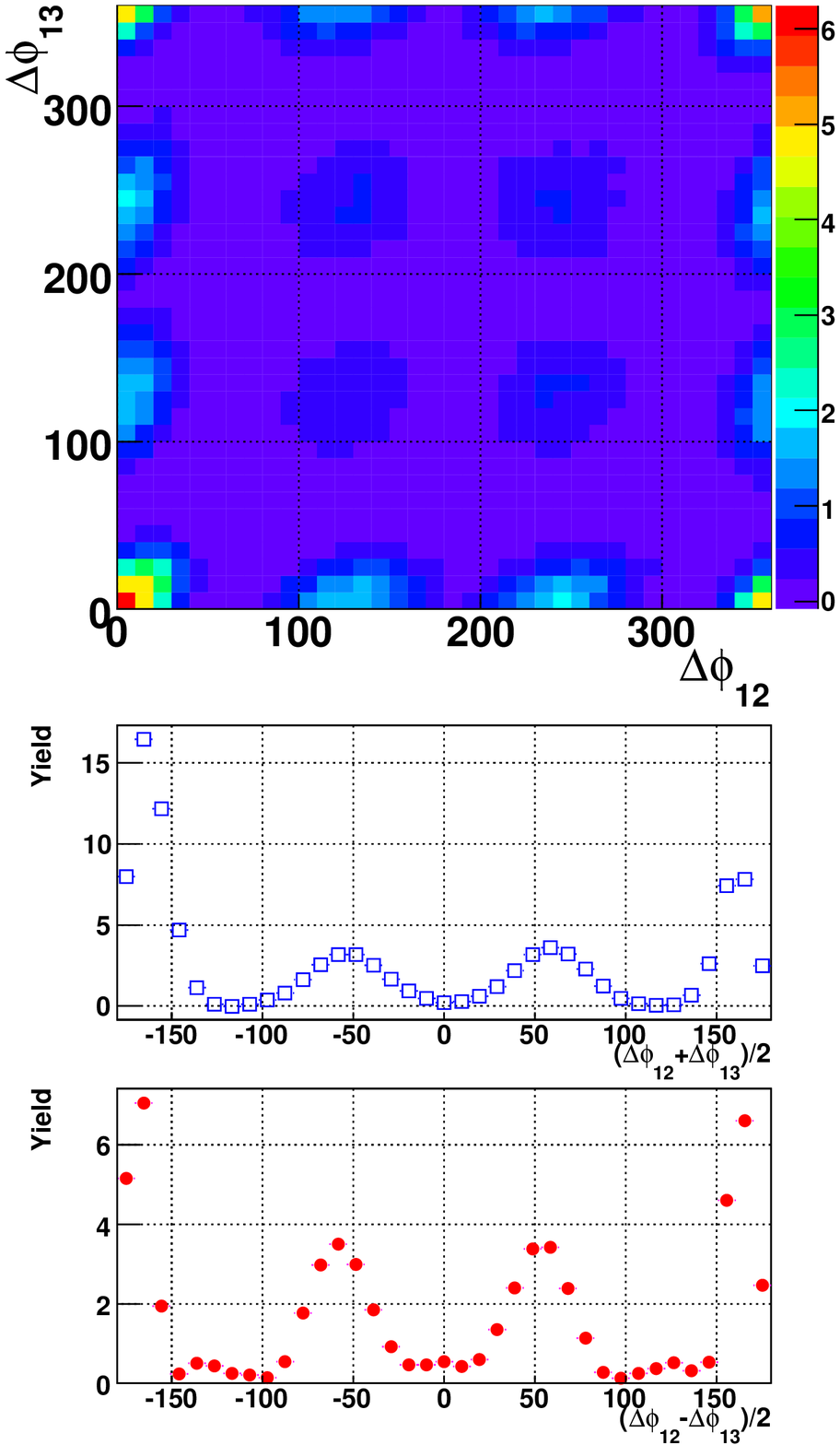} \end{center}
\caption[]{Top: Illustration of conical signal expected based on monte carlos simulation described in the text. Middle/Bottom: 40 degree slice projections along the main and alternate diagonals.}
\label{fig2}
\end{minipage}
\begin{minipage}{0.5\linewidth}
\begin{center}\includegraphics[width=3.1in,height=3.8in]{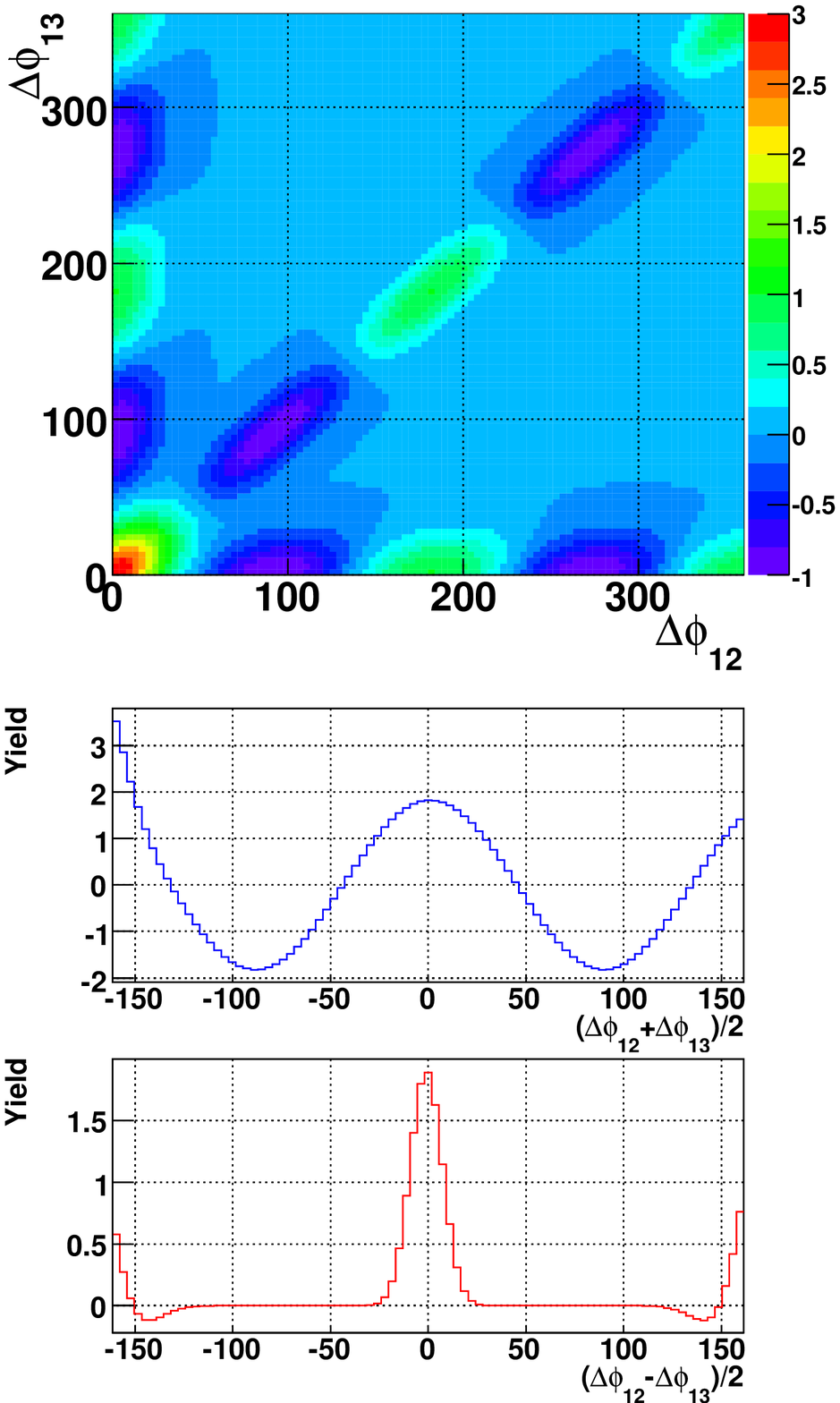} \end{center}
\caption[]{Top: Jet-Flow 3-cumulant component obtained with Eq. 6 for $a_i = a_j = 0.5$ plotted in arbitrary units. Middle/Bottom: Projections along the main and alternate diagonals.}
\label{fig3}
\end{minipage}
}
\end{figure}

\section{Summary}

We presented a new technique based on three-particle cumulants for the study of jet structure and searches for conical emission in heavy ion collisions at RHIC and LHC. While the cumulant technique was introduced for three-particle azimuthal correlations, it can be trivially extended to include correlations in rapidity space as well.  We showed, with simple models of jet, flow, and conical emission, that it should be possible to distinguish these  processes using three-particle correlations.  Additional effects due to momentum conservation discussed in \cite{Borghini06} should also have distinct features in three-particle cumulants. We thus believe that searches for conical emission with this type of observable should yield a definite and relatively unambiguous answer after proper treatment of flow, jet-flow background correlations, and momentum conservation effects.

{\bf Acknowledgments}
This work was supported, in part, by U.S. DOE Grant No. DE-FG02-92ER40713.

\end{document}